\documentclass[12pt]{article}

\usepackage{epsfig}
\def\sect #1{\setcounter{equation}{0}}
\def\tc{t_{sc}}
\def\tf{t_f}
\def\diff#1#2{\frac{\partial #1}{\partial #2}}

\begin{document}

\title{\large \bf GRAVITATIONAL COLLAPSE
OF FLUID BODIES \\ AND COSMIC CENSORSHIP: ANALYTIC INSIGHTS}

\author{S. Jhingan\thanks{E-mail : sanjhi@mate.polimi.it} \, and G.
Magli\thanks{E-mail : magli@mate.polimi.it}\\
{\it Dipartimento di Matematica, Politecnico di Milano}\\
{\it Piazza Leonardo da Vinci 32, 20133 Milano, Italy}}
\date{}
\maketitle

\begin{abstract}
The present analytical understanding on 
the nature of the singularities which
form at the endstate of gravitational collapse of massive fluid bodies 
(``stars") is reviewed.
Special emphasis is devoted to the issue of physical reasonability of
the models.
\end{abstract}

\section{Introduction}

The investigation on the ``final fate'' of
gravitational collapse of initially regular distributions
of matter is one of the most active field of research
in contemporary general relativity. It is, indeed, known that under fairly 
general hypothesis solutions of the Einstein 
equations with ``physically reasonable" matter can develop into 
singularities~\cite{Hawking-Ellis}. The key problem that still
remains unsolved is the {\it nature} of such singularities. 
The main open issue is whether the singularities, which
arise as the end point of collapse, can actually be observed.

Roger Penrose~\cite{Penrose69},
was the first to propose the idea, known as cosmic censorship conjecture:
{\it does there exist a cosmic censor who forbids the occurrence of naked
singularities, clothing each one in an absolute event horizon?}
This conjecture can be formulated in a ``strong'' sense 
(in a ``reasonable" spacetime we cannot have a naked singularity) 
or in a weak sense
(even if such singularities occur they are safely hidden behind an event 
horizon, and therefore cannot communicate with far-away observers).  
Since Penrose's proposal
there have been various attempts to prove the 
conjecture (see~\cite{Global} and references therein).
Unfortunately, no such attempts have been successful so far.
As a consequence, the research in this field
turned to more tractable objectives.
In particular, one would like to understand
what happens in simple systems, like spherically symmetric  
ones (interestingly enough, even this apparently innocuous problem is far
from being completely solved, although, as we shall see, a general
pattern does seem to arise).

Our aim here is to overview only
models which have a clear physical interpretation.
Therefore, we shall require satisfaction of the weak energy condition
as well as  existence of a singularity free initial data surface.
Moreover, we shall  take into consideration solutions
of the Einstein field equations which are physically meaningful in terms 
of a (phenomenological) equation of state of the matter.
In this respect it is worth mentioning that material schemes having
a well defined {\it microscopical} interpretation,
like the Vlasov-Einstein system, would be closer
to such a requisite~\cite{Rendallrev}. 
However, very little is known
on the dynamics of such models (a numerical investigation 
has been recently carried out~\cite{rrs}).

In presence of excellent general reviews on gravitational collapse
and cosmic censorship~\cite{Israel1, Eardley1, Clarke1, Waldrev, Joshirev, 
TPrev}, we have focused our attention 
on a specific issue namely, the investigation of
analytical models describing the gravitational collapse of massive stars.
Therefore we are not going to address here many related important
topics. These include Vaidya spacetimes~\cite{Global}, 
radiation shells (see~\cite{Waldrev} and references therein),
gravitational collapse of scalar fields~\cite{Ch94, Ch99}, 
critical behaviour in numerical relativity~\cite{Gundlacrev}, 
stability of Cauchy horizon in Reissner-Nordstr\"om 
spacetimes~\cite{Ori-Burko}, the Hoop conjecture~\cite{KThorne, Lemos99} 
among others~\cite{Israel1, Joshirev, Israel2, Israel3}.

\section{Einstein equations for spherical collapse}

What is known {\it analytically} in gravitational collapse
is essentially restricted to spherical symmetry (one exception is given
by the Szekeres ``quasi-spherical" spacetimes~\cite{Szekeres}, 
but the results~\cite{Krolak, Deshingkar} are very similar to those holding 
in Tolman-Bondi models -- See Section 4.2.4).
Therefore we discuss the mathematical structure
of the Einstein field equations describing collapse 
of a deformable body only in the spherically symmetric case. 
For perfect fluids
this structure is well known~\cite{Misner};
we present here a more general case which 
takes into account anisotropic materials as well~\cite{Magli1}.
Also, we  consider only non-dissipative processes
since very little is known in the dissipative case~\cite{Herrera-Martinez}.

The general, spherically symmetric, non-static
line element in comoving coordinates
$t,r,\theta,\varphi$  
can be written in terms of three functions $\nu , \eta , Y$
of $r$ and $t$:
\begin{equation}
ds^2= -e^{2\nu}dt^2 + {\eta}^{-1} dr^2 + Y^2 (d\theta^2 + \sin^2\theta
d\varphi^2)\ .\label{metric}
\end{equation}

Throughout this paper, we shall assume that the 
collapsing body is ``materially spherically symmetric''
in the sense that all the physical ``observables''
do not depend on angles.
The matter density of the material (baryon number density)
is then given by
\begin{equation}
\rho =\rho_0 (r) Y^{-2}\sqrt{\eta} \ ,\label{rho}
\end{equation}
where $\rho_0 (r)$ is an arbitrary (positive) function.
As in any theory of continuous media,
to choose a specific material one has to specify
the {\it internal energy}  $\epsilon$.
This function depends on the parameters characterizing
the state of strain of the body (for a recent review
of relativistic elasticity theory see~\cite{Kijowski}). 
It can be shown that 
such parameters can be identified with
$Y$ and $\eta$ in the comoving frame (in other words, any deformation
is described ``gravitationally'').
Therefore, we introduce 
as {\it equation of state} of the material
a positive function 
\[
\epsilon = \psi  (r,Y,\eta) \ .
\]
Here the explicit dependence on $r$ takes into account 
possible inhomogeneities.

The energy-momentum tensor can be readily calculated and
the result is a diagonal tensor of the form
$T^\mu_\nu ={\rm diag} (-\epsilon , \Sigma , \Pi ,\Pi )$.
Here the stress-strain relations (i.e. the 
relations giving the radial stress $\Sigma$ and the tangential stress
$\Pi$ in terms of the 
constitutive function)
are given by
\begin{equation}
\Sigma =2\eta \diff{\psi }{\eta}
-\psi \ ,\label{state1} 
\end{equation}
\begin{equation}
\Pi =-\frac 12 Y\diff\psi {Y} -\psi \ .\label{state2}
\end{equation}
We shall always use the word stress rather than pressure since
both $\Sigma$ and $\Pi$ can be in principle negative (tensions)
without violating the energy conditions.

Different materials correspond to different choices of the function
$\psi$.
It is, however, worth mentioning that the material scheme
most widely used in astrophysical applications 
is the perfect fluid, which can be characterized 
as a material whose function of state depends on the number
density only ($\epsilon = \tilde \psi (\rho)$).
In this case both stresses coincide: 
\[
\Pi =\Sigma =\rho \frac{d\tilde \psi
}{d\rho} -\tilde \psi :=p \ ,
\]
where $p$ is the isotropic pressure.
Two particular cases are worth mentioning in this scheme.
One is that of linear pressure-density relation
($p=k\epsilon$); the 
corresponding equation of state is  $\tilde \psi (\rho) =\rho(1+A\rho^{k})$
where $A$ is a constant.
The other one is the {\it dust model} for which $p=0$.
In this case 
the energy is distributed proportionally to the mass
($\tilde\psi(\rho)=\rho$).

As soon as one allows anisotropy to occur,  other interesting models 
appear (see~\cite{Herrera-Santos} for a review on the role of anisotropy 
in relativistic astrophysics).
Recently, a particular anisotropic model has been 
singled out~\cite{Magli1, Magli2}
(for previous investigations on this kind of models
see references in~\cite{Magli2}).
In this model, one assumes that the radial stress
identically vanishes. The key role is played by the equation 
(\ref{state1}) which shows that the dependence of $\psi$ on $\eta$ 
must be a multiplicative dependence from $\sqrt{\eta}$ only.
Therefore, materials with vanishing radial stresses can be characterized,
using equation (\ref{rho}),  
{\it via} equations of state of the form
\begin{equation}
\psi (r,\eta ,Y)=\rho h(r,Y)  ,
\label{tang}
\end{equation}
where $h$ is a positive, but otherwise arbitrary, function.

Once an equation of state has been chosen,
the Einstein field equations become a closed system;
in spherical symmetry there are three independent equations
for the three variables  $\nu $, $\eta$ and $Y$.
It has proven, however, to be very useful to write the field equations
as a system of {\it four} differential equations.
This is done by introducing
the {\it mass function}~\cite{Misner, Lemaitre}
defined as 
\begin{equation}
m(r,t)=\frac Y2 \left(
1-Y'^2\eta
+\dot Y^2e^{-2\nu}\right) \ ,
\label{massa}
\end{equation}
where a dash and a dot denote derivatives with respect to $r$
and $t$ respectively.
The mass function is arbitrary (positive) 
and allows us to write the field equations in
the following form (four compatible equations for three variables)
\begin{equation}
m' =4\pi \epsilon Y^2 Y'\ ,\label{efe1}
\end{equation}
\begin{equation}
\dot m =-4\pi \Sigma Y^2 \dot Y\ \ ,\label{efe2}
\end{equation}
\begin{equation}
Y'\dot\eta =-2\eta (\dot Y' - \dot Y\nu') \ ,\label{efe3}
\end{equation}
\begin{equation}
\Sigma'
= -(\epsilon+\Sigma )\nu' -2 (\Sigma -\Pi) (Y'/Y)
\ .
\label{efe4}
\end{equation}

\section{Physical reasonability and initial data}

It is easy to produce new ``solutions" of the Einstein field equations in
``matter". Indeed, just pick up a metric at will and claim that this is a
``solution" referring to the {\it calculated} energy momentum tensor.
Of course, what one has to do to remove the quotation marks from the above
statements is to check the physical reasonability of the results.

First of all, the weak energy condition 
must be imposed on the energy momentum tensor.
This condition requires $T_{\mu \nu} u^\mu u^\nu \geq 0$ for any
non spacelike $u^\mu$ and implies, besides positivity of 
$\epsilon$, non-negativeness of $\epsilon +\Sigma $ and $\epsilon +\Pi$.
Due to equations (\ref{state1}) and (\ref{state2}), such conditions 
are equivalent to
differential inequalities on the function $\psi$, namely,
$\diff\psi\eta \geq 0$ and $\diff\psi{Y}\leq 0$.

Imposing the weak energy condition 
{\it per se} does not assure physical reasonability,
since there is no guarantee that the energy momentum tensor 
can be deduced from a field theoretic description of matter. 
What is needed is the satisfaction of a suitable
equation of state. 
We shall require, in addition, the equation of state to be locally 
stable~\cite{Magli1} (this last
requirement could be relaxed in presence of rotationally-induced stress). 
The local stability condition requires the (local) equilibrium state of the
material to be unstrained. In the ``comoving picture''
this amounts to say that the energy must have an absolute minimum at the 
flat-space values of the metric.

Let us collect the functions describing admissible
equations of state in spherical symmetry in a set
\[
{\bf \Psi} =\left\{
\psi \in C^2({\rm R_+^3}, {\rm R_+}):
\psi (r,r,1)=\min \psi(r,Y,\eta),
~\diff\psi\eta \geq 0, \diff\psi{Y}\leq 0
\right\} \ .
\]
We shall always assume that the value  
of $\psi$ at the minimum (which in general can be a function of $r$)
has been rescaled to unity (this can be done without loss of generality).

A solution of the Einstein field equations describes the collapse 
of an initially regular distribution of matter
only if the spacetime admits a 
spacelike hypersurface ($t=0$ say) which carries {\it regular} initial data.
This means that the metric, its inverse, 
and the second fundamental form all have to be continuous at $t=0$.

On the initial hypersurface we 
use the scaling freedom of the $r$ coordinate 
to set $Y(r,0)=r$. We call a set of initial data {\it complete} 
if it is minimal, in the sense that no part of its content can be gauged away
by a coordinate transformation. We will now prove that
a complete set of initial data for equations (\ref{efe1})-(\ref{efe4}),
at fixed equation of state, is composed by a pair of functions.
Physically, such functions describe the initial
distribution of energy density $\epsilon_0=\epsilon(r,0)$
and of velocity $V_0=e^{-\nu (r,0)} \dot Y(r,0)$.
It is sometimes convenient to parameterize
these two distributions in terms of two other functions 
$\{F(r), f(r)\}$ 
where $F(r) = m(r,0)$ is the initial distribution of mass and
$f(r)$ is called ``energy function".
The relationship between the two sets $\{F,f\}$
and $\{\epsilon_0,V_0\}$ is given by the following formulae
\begin{equation}
F'=4\pi r^2 \epsilon_0\ ,\ \ f=
V_0^2-2F/r\ .\label{fprimo}
\end{equation} 
The function $F$ has to be non-negative with
$ \epsilon_0(0) = \lim\limits_{r\to 0}F(r)/r^3$ finite and non-vanishing,
while $f$ has to be greater than $-1$ to preserve the signature of the 
metric (see equation (\ref{omega}) below) with $\lim\limits_{r\to 0}f(r)=0$.

The proof that only two arbitrary functions 
of $r$ are a complete set of initial data at fixed equation of state
is implicitly contained in many papers on spherical collapse
(see e.g.~\cite{Misner, JoshiCQG}).
It seems, however, that a complete proof has never been 
published in details, so we take this occasion to give it.
We denote by a subscript, the initial value of
each quantity appearing in the Einstein field equations. We
know $F=m_0$, $V_0=e^{-\nu_0}\dot Y_0$.
Since $Y_0=r$ and $\psi$ is a known
function of $r,Y$ and $\eta$, its initial value $\psi_0$
is also known - as well
as the initial values of the stresses due to formulae 
(\ref{state1}) and (\ref{state2}) - as a function of $r$ and $\eta_0$
($\psi_0=\psi (r,r,\eta_0)$).
Evaluation of equation (\ref{efe1}) at $t=0$ now gives
$\psi_0 =F'/(4\pi r^2)$, from which 
the value of $\eta_0$ can be extracted algebraically. 
At this point, the remaining three field equations 
can be used to evaluate the remaining data, i.e., $\dot m_0$, $\nu_0$ and
$\dot \eta_0$:

\begin{equation}
\dot m_0 = -4\pi {\Sigma}_0 r^2 V_0\ , \
{\nu'}_0 = -\frac{\Sigma'_0}{\psi_0 +\Sigma_0}
-2 \frac{\Sigma_0 - \Pi_0}{r(\psi_0 +\Sigma_0)}\ , \
\dot \eta_0
= -2\eta_0 e^{\nu_0} {V'}_0\ .
\end{equation}
This completes the proof.

In what follows,  we consider only solutions  which can be
interpreted as models of collapsing stars,
i.e. isolated objects rather than ``universes''.
This is possible only if the metric matches smoothly
with the Schwarzschild vacuum solution
(the matching between two metrics is smooth
if both the first and the second
fundamental form are continuous on the matching surface).

\section{Classification and nature of singularities}

To understand the collapsing
scenarios as well as the nature of the singularities, one 
would like to analyse exact solutions of
the Einstein field equations.
However, it goes without saying that the non-linearity
of such equations makes them essentially untractable, even in spherical 
symmetry, without further simplifying hypotheses (like e.g.
vanishing shear or acceleration, see~\cite{Kramer, Krasinski}).
In view of such difficulties,
one would like to extract information from the Einstein equations 
without solving them completely.

The mathematical structure of spherical collapse
discussed in the previous Section
shows that there is a one-to-one correspondence between
solutions and choices 
of triplets ($s$, say)
of functions $s=\{F(r), f(r), \psi (r,\eta,Y)\}$
(we are, of course, identifying solutions {\it modulo}
gauge transformations).
We denote the set of solutions parameterized in this way 
by ${\cal S}$.
Since in this parameterization we have already taken
into account regularity as well as physical admissibility,
the whole physical content of the cosmic censorship problem can
be translated in the mathematical terms of predicting the endstate
of any choice of $s \in {\cal S}$.

\subsection{Non-singular solutions.}

The existence
of non-singular, non-static solutions of the Einstein field equations
is not forbidden by the singularity theorems
(we refer the reader to~\cite{Senovilla} for a recent review on singularity
theorems and further references) and
a gravitational collapse can lead to a bouncing back,
at a finite area radius, without singularity formation.
This phenomenon can be recursive, producing an eternally oscillating, 
globally regular solution~\cite{Bonnor1, Thompson68, Bondi69}.
Speaking very roughly, 
one can prepare
regular initial data in such a way that the region of possible
trapped surface formation is disconnected 
from the data. In this way the remaining hypothesis of the singularity 
theorems can be satisfied 
without singularity formation (see also~\cite{Magli1}).
  
Much more exotic than the oscillating solutions, 
are the globally regular solutions
that describe non-singular blackholes~\cite{Senovilla}.
These are matter objects ``sitting'' inside their Schwarz\-schild radii. 
Their finite extension replaces the central singularity
with a matter-filled, non singular region.
In this case trapped surfaces are obviously
present and therefore
the strong energy condition must be violated (a simple argument
shows that also the dominant energy condition is necessarily violated).
In any case, such strange objects are not ruled out by general relativity
as far as only the weak energy condition and an equation of state
are required~\cite{Mars, Ayon, Magli3}.
However, as far as we are aware, 
it is not known if a fully dynamic solution 
exists which could eventually lead to this exotic end state.

\subsection{Singular solutions}

We now move to the case in which a singularity is formed
in the future of a regular initial data set.
The singularities of spherically symmetric 
matter filled spacetimes can be recognized from divergence of the 
energy density and curvature scalars, like e.g. the 
Kretschmann scalar $R^{\mu\nu\rho\sigma}R_{\mu\nu\rho\sigma}$.
Essentially, these singularities can be of two kinds.
We shall call {\it shell crossing}
singularities those at which $Y'$ vanishes ($Y\neq0$), 
and {\it shell focusing} singularities those
at which $Y$ vanishes.
To the two kind of singularities 
correspond two curves $\tc (r)$ and $\tf (r)$
in the $r$-$t$ plane, defined by
$Y'(r,\tc(r))=0$ and by $Y(r,\tf (r))=0$ respectively.
Physically, the shell crossing curve gives
the time at which two neighbouring shells of matter
intersect each other, while the shell focusing curve identifies at which
time the shell labeled $r$ ``crushes to zero size''.
Of these two kinds of singularities, the one of physical interest
is obviously that occurring first in the sense that, at fixed $r$,
one has $\tf >\tc$ or vice versa.
It would be very interesting, therefore, to carry
out a study of the field equations in order to obtain conditions for 
shell crossing avoidance in spherical spacetimes.
This study has been up to now carried out only for dust 
spacetimes~\cite{Hellaby, Newman, Jhingan}.

From the point of view of censorship,
the nature of a singularity in an asymptotically flat, 
initially regular spacetime
can be one of the following.
First of all, a singularity can be {\it spacelike}, 
like e.g. the Schwarzschild
singularity or the singularity occurring
at the endstate of the collapse of a Oppenheimer--Snyder 
dust cloud (see Section 4.2.2 below). 
These singularities lie in the future
of all possible observers and therefore are strongly censored 
(i.e. allowed by strong cosmic censorship).
If a singularity is not strongly censored, then it is naked,
i.e. visible to some observer.
However, two different cases can occur, namely 
the singularity can be 
{\it locally} or
{\it globally} naked.
A singularity
is locally naked if light signals can emerge 
from it but fall back
without reaching any asymptotic observer.
A singularity of this kind 
will be visible only to  
observers who have crossed 
the horizon; therefore,
the weak cosmic censorship holds for such endstates.
An example of a spacetime containing a locally naked singularity is provided
by the Kerr spacetime with mass greater than the angular
momentum per unit mass (or by the Reissner--Nordstr\"om spacetime
with mass greater than charge). 
Finally,
a singularity
is globally naked if light-rays emerging from it
can reach an asymptotic observer. 

\subsubsection{Shell crossing singularities.}

The first explicit example showing formation of a naked singularity 
was found as a shell crossing singularity in a spherical dust 
cloud~\cite{Yodzis1}. It can be shown that these singularities are
timelike and are always locally naked.

Some definitions have been proposed to put singularities ``in the
order of increasing seriousness''~\cite{Clarke-Krolak}.
Essentially, what is done is to 
check the behaviour of the invariants
of the Riemann tensor in the approach to the singularity.
According to such criteria,
the shell crossing singularities 
turn out to be ``weak'' at least as compared with the shell focusing 
singularities.
This ``weakness' is considered by some authors as an hint of a possible
extension of the spacetime~\cite{Clarkebook}.
However, in spite of their ``weakness'',
there is at the moment no available general proof of extendibility
of spacetimes through a shell cross
(although some encouraging results exist, see~\cite{ODonell}).
The unique exception is a paper by Papapetrou and Hamoui~\cite{Hamoui}.
In this paper the authors claim to have explicitly found
the extension in the case of ``degenerate'' shell crossing singularities,
i.e. when the curve $\tc (r)$ degenerates to an ``instant of time''
$\tc =T= {\rm const.}$
In this case, it is easy to check that the crossing happens 
at a ``point'' $r={\rm const.}$ rather than at a ``3-space''
and this is the key to their treatment. 
However, some results of this paper are unclear from the
physical point of view. 

What is actually available 
is only a {\it continuous} extension of shell crossing
singularities exists in the dust case~\cite{Newman}.
We shall show this in a slightly more general case.
Integrating equation (\ref{efe3}) formally with respect to time we can write
\begin{equation}
\eta^{-1}=\frac {Y'^2}{1+f}\Omega^2 \ ,\label{omega}
\end{equation}
where $f$ is the energy function
and $\Omega (r,t)=
{\rm exp}\left(-\int_0^t \frac{\nu'\dot Y}{Y'} d\tilde t\right)$.
Changing variable from $r$ to $Y=Y(r,t)$ one gets the metric
\[
ds^2 =\frac{\Omega^2}{1+f}
\left[\left(\dot Y^2 -(1+f)\Omega^{-2}e^{2\nu}\right)
dt^2 -2\dot Y dYdt +dY^2\right]+Y^2
(d\theta^2 + \sin^2\theta
d\varphi^2)
\]
If $\Omega$ 
is finite and non-vanishing 
at $t=\tc (r)$,
the above metric is continuous with
continuous inverse at such surface.
This happens if $\nu'$ vanishes (dust case)
or, more generally, if $\nu' $ goes to zero at least as $Y'$
at the shell cross (this happens, for instance, in the case
of vanishing radial stresses).

In a recent paper~\cite{Lun}, Szekeres and Lun have shown
that there exist a system of coordinates in which the metric is of class 
$C^1$ as $t\to \tc^-$. However, again, this result {\it per se}
does not show extendibility of the spacetime (see 
also~\cite{Iyer}).

\subsubsection{Non-central shell focusing singularities }

It is important to distinguish the {\it central}
shell focusing singularity, i.e. that occurring at $r=0$, 
from the other focusing singularities, since
in many cases it is easy to prove that
non central singularities are 
censored.

A necessary condition for the visibility of a ``point" $r$
is that the condition $1-\frac{2m(r,t)}{Y(r,t)} > 0$, implying absence of 
trapped
surfaces (see next Sub--section),
is satisfied. Since the apparent horizon is the 
boundary of the region containing
trapped surfaces, the above condition implies that the time of formation
of the apparent horizon  $t_{ah}(r)$, defined as $2m(r,t_{ah}(r)) = 
Y(r,t_{ah}(r))$, must not be {\it before}  $\tf (r)$, i.e. 
$t_{ah}(r) \geq  \tf (r)$.
Now suppose $m(r,\tf (r))$ to be different from zero
(as we have seen, this is not the case at the central singularity
where $m$ has to vanish as $r^3$).
Then $1 - 2m(r,t)/Y(r,t)$ goes to minus infinity as $t$ tends to $\tf (r)$
so that the singularity is covered.
This shows that all the naked shell focusing singularities
are necessarily massless, in the sense that $m$ has to vanish 
there~\cite{JoshiCMP}.
It follows that any non central singularity will certainly be censored
if the mass is an increasing function with respect to $t$.
Now equation (\ref{efe2}) gives $\dot m=-4\pi \Sigma Y^2 \dot Y$
and, since $\dot Y<0$ during collapse, we conclude
that non central singularities are always covered if the radial stress
$\Sigma$ is non-negative~\cite{Cooperstock} 
(in particular, all non-central singularities occurring
in dust as well as in models with only tangential stress
are covered, since the mass does not depend on time in this case).  
In the presence of radial tensions, the question is still open.
It is known that a perfect fluid with $p=k\epsilon$ 
exhibits naked non-central singularities if $k<-1/3$~\cite{Cooperstock}.  

\subsubsection{Central singularities: the root equation }

The first explicit examples of the formation of a naked shell focusing 
singularity were provided by 
Eardley~\cite{Eardley74}, Eardley and Smarr~\cite{Eardley79} and by
Christodoulou~\cite{Christodoulou1}.
Since then the techniques to study
the nature of the singularities in spherically symmetric spacetimes
have been developed by many authors (see references in ~\cite{Jhingan}) 
and finally settled up by Dwivedi and Joshi~\cite{JoshiPRD}.

The key idea is the following:
if the singularity is visible, at least locally,
there must exist light signals coming out from it.
Therefore, by investigating the behaviour of radial null geodesics
near the singularity, one can try to find out if outgoing null curves 
meet the singularity in the past.
On such radial null geodesics,
the derivative of $Y(r,t)$ reads
\[
\frac {dY}{dr} =Y' +\eta^{-1/2}\dot Y e^{-\nu}\ .
\]
Using (\ref{massa}) and (\ref{omega})
in the above equation we obtain
\begin{equation}
\frac {dY}{dr} =Y'\left[
1-\sqrt{1+\frac {\Omega^2}{1+f}
\left(\frac {2m}Y-1\right)} \right]\ ,\label{root}
\end{equation}
where $Y'$ has to be understood as a known function of $Y$ and $r$.
If the singularity is naked, equation (\ref{root}) must have at least one 
solution with definite, outgoing tangent at $r=0$, i.e. a solution of
the kind $Y=X_0r^\alpha$ where $\alpha>1$ and $X_0$ is a positive 
constant. Clearly, this behaviour 
is possible only if the necessary condition 
$1-{2m}/Y>0$ is satisfied. Indeed 
$Y'$ is equal to one, and therefore
positive, on the initial data surface.
If no shell crossing occurs, it
remains positive, so that the right hand side 
of equation (\ref{root}) cannot remain positive 
if ${2m}/Y-1$ changes sign.

Once the necessary condition is satisfied, one has to check
if both $X_0$ and $\alpha$ exist such that the solution of
equation (\ref{root}) is of the specified form near the singularity.
On using L'Hospital rule we have 
\begin{equation}
X_0=\lim_{r\to 0}
\left(\frac1{\alpha r^{\alpha-1}}\frac{dY}{dr}\right)_{Y=X_0r^\alpha} .
\label{root1}
\end{equation}
Using again (\ref{root}), this equation becomes an {\it algebraic}
relation for $X_0$ at fixed $\alpha$. If a positive
definite $X_0$ exists  the singularity is naked.

\subsubsection{The dust case}

The general exact solution of the Einstein field equations 
is known in the most simple case of vanishing pressure 
(dust)~\cite{Lemaitre, Tolman, Bondi}. In this case,
from equation (\ref{efe4}), 
one gets $\nu =0$ (more precisely, $e^{\nu}$ is an arbitrary function
of $t$ only which can be rescaled to unity without loss of generality).
Then $\Omega = 0$ and it follows from equation (\ref{omega}) that 
$\eta^{-1}=Y'^2/(1+f)$.
The mass is constant in time ($m=F(r)$) due to equation (\ref{efe2})
with $\Sigma=0$.
Therefore (\ref{massa}) can be written as 
a Kepler-like equation
($\dot Y^2 =f+2F/Y$), which is integrable
in parametric form for $f\neq 0$ and
in closed form for $f=0$. Finally, the density can be read off
from (\ref{efe1}) as $\epsilon =F'/(4\pi Y^2 Y')$.

A great effort has been paid to understand the nature of the central 
singularity in this 
solution~\cite{Newman, Jhingan, Eardley79, Christodoulou1, JoshiPRD, Oppen, 
Gorini, Grillo, Lake92, TPSingh} and we now 
know the complete spectrum of possible endstates of the dust evolution
in dependence of the initial data.
We recall here what happens in the case of marginally bound solutions 
($f=0$)~\cite{TPSingh}
since it is sufficiently general to illustrate a tendency and 
simple enough to be recalled in a few lines.

For marginally bound dust, the solutions $s = \{F,1,1\}$ can be uniquely
characterized by the expansion of the function $F(r)$ at $r=0$
or, and that is the same, by the expansion of $\epsilon_0=F'/4\pi r^2$.
Using this expansion in the root equation, it is not difficult to check
the following results:
\begin{itemize}\item
If the first non--vanishing term 
corresponds to $n=1$ or $n=2$ equation (\ref{root1}) always has a real positive 
root: the singularity is naked;
\item
If the first non--vanishing term is $n=3$ the root equation 
reads 
\begin{equation}
2x^4 + x^3 + \xi x  - \xi = 0 \ ,
\label{DUQUART}
\end{equation}
where $x^2=X$ and $\xi = F_3/{F_0}^{5/2}$. From the theory of quartic one 
can show that this 
equation admits a real positive root if 
$\xi <\xi_c =-(26+15\sqrt{3})/2$. Therefore, $\xi_c$ is a critical parameter:
at $\xi =\xi_c$ a ``phase transition'' occurs and
the endstate of collapse turns from a naked singularity
to a blackhole.
\item
If $n>3$ the limit in 
equation (\ref{root})
diverges: the singularity is covered.
In particular, this case contains the solution
first discovered by Oppenheimer and Snyder~\cite{Oppen}
describing a homogeneous dust cloud.
\end{itemize}

The naked singularities mentioned above are
{\it locally} naked. 
It can, however, be shown that if locally naked singularities 
occur in dust spacetimes, then spacetimes containing globally naked 
singularities
can be build up from these by matching procedures.
The Penrose diagrams 
corresponding to the three different cases are shown in Figure 1.

The above results can be extended to the general case of 
collapsing dust clouds, so that
the final fate of the dust solutions
$s = \{F,f,1\}$ is completely known.
The final fate depends on a parameter which is a combination
of coefficients of the expansions 
of $F$ and $f$ near $r=0$, and  
a structure similar to that of marginally bound collapse
arises (see~\cite{Jhingan} for details).

\begin{figure}
\centerline{\mbox{\epsfig{file=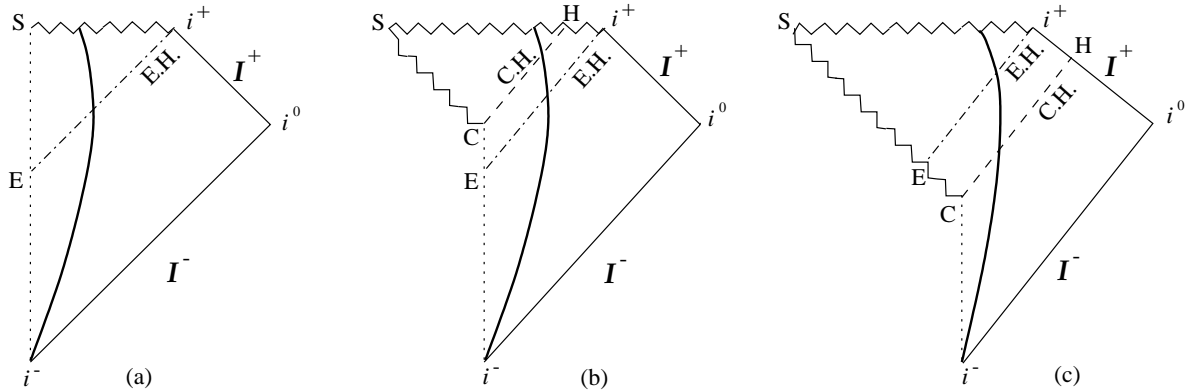}}}
\caption{\small{Penrose diagrams for collapsing dust clouds.
 The dotted line represents the center of the star, and the bold
jagged line
represents the singularity.
The continuous line connecting ${\it i}^{-}$ to the singularity
represents the boundary of the collapsing cloud.
Three different situations can occur, depending on the choice
of initial data:
(a) ``generalized''
Oppenheimer-Snyder collapse: the singularity is covered.
(b) locally naked singularity (c) globally naked
singularity.}}
\end{figure}

\subsubsection{Vanishing radial stresses}

Recently, the general solution for spherically symmetric dust
has been extended to the case in which 
only the radial stress vanishes~\cite{Magli1, Magli2}.
The solution can be reduced to quadratures
using a system of coordinates first introduced by Ori~\cite{Ori} 
for charged dust.
One of the new coordinate is the mass $m$ which 
is constant in time (due to (\ref{efe2}) with $\Sigma=0$),
the other coordinate is the ``area radius'' $Y$.
In such coordinates the metric reads

\begin{equation}
ds^2= -\Gamma^2\left(1-\frac {2m}Y\right)dm^2 
+2 \sqrt{1+f}\frac{\Gamma}{hu} dY dm -\frac 1{u^2}dY^2
+Y^2 (d\theta^2 + \sin^2\theta
d\varphi^2)\ ,
\end{equation}
where
\begin{equation}
u = -\sqrt{-1 + \frac{2m}{Y} + \frac{1+f}{h^2}} \ , \,
\Gamma
= 
g(m) +\int 
\frac h{u^2\sqrt{1+f}} \frac{\partial u}{\partial m} dY
\ ,
\label{m10}
\end{equation}
and the function $g(m)$ is arbitrary.

The problem of understanding the nature of the singularities for such
solutions is essentially still open.
It is, indeed, possible to write the root equation in 
explicit form, but this equation contains a sort of ``non-locality''
due to the integral entering in the definition of $\Gamma$.
As a result only a few special cases have been analysed so 
far~\cite{Magli2, WittenSingh, Barve}.

Among the solutions with tangential stresses 
a particularly interesting one is
the {\it Einstein cluster}~\cite{Einstein, Datta, HBondi}.
This is a spherically symmetric
cluster of rotating particles. 
The motion of the particles is sustained by the angular momentum $L$
whose average effect is to introduce a non vanishing tangential stress
in the energy-momentum tensor.
The corresponding equation of state has the form
$h(m,Y)=\sqrt{1+L^2(m)/Y^2}$.
Therefore, a solution is uniquely identified by
the choice of three arbitrary functions of $m$ only, namely $F,f$ and $L^2$
(for $L=0$ one recovers dust).
It turns out that the final state ``at fixed dust
background'' (i.e. for fixed $F,f$) depends on the expansion
of $L^2$ near $m=0$ ($L\approx \beta m^y$, 
say)~\cite{Haradaprd, JhinganMagli}.
Considering, for simplicity, the marginally bound case,
one finds that 
for $4/3 <y <2$ either the singularity does not form
(the system bounces back) or a blackhole is formed.
The threshold of naked singularity formation lies at $y=2$,
where a $2$-parameter structure very similar to that of dust occurs. 
At $y=7/3$ a sort of transition takes place and the evolution of the 
model is such that only the critical branch is changed, un-covering a part of 
the blackhole region in the corresponding dust spacetime; the non-critical
branch is the same as in dust spacetimes. 
Finally for $y>7/3$ the evolution always leads to the same end state
of the corresponding dust solution.

\subsubsection{Self-similar collapse}

A spherically symmetric spacetime is self-similar if it admits an
homothetic vector $\xi$, i.e. a vector satisfying 
${\cal L}_\xi g_{\mu\nu} = 2 g_{\mu\nu}$.
In the comoving frame
the dimensionless variables $\nu$, $\lambda$ and $Y/r$ depend only
on the ``similarity variable'' $z:=r/t$,
and the Einstein field equations become {\it ordinary}
differential equations
(we refer the reader to~\cite{Carr98} for a complete
treatment of self-similar solutions).
Being governed by ordinary differential equations,
self-similar spherical collapse can
be analyzed with the powerful techniques of dynamical
systems theory~\cite{Bogoyavlenski}.
The analysis of the singularities forming in self-similar spacetimes
has been done by many authors for different equations of state,
like dust~\cite{Eardley74, Eardley79, Wagh88, OriPiran, Lemos91, Lemos92},
barotropic perfect fluids~\cite{OriPiran, Ori87, Joshi92},
radiation (Vaidya) shells~\cite{Kuroda, Papapetrou85, Lake86} 
and in general cases~\cite{Lake89, Lake90}.

The picture arising resembles the dust case
in the sense that both naked singularities and blackholes can
form depending on the values of the parameters
characterizing the solution.
A thoroughly review of these and other features of self-similar solutions
can be found in~\cite{Carr97};
here we limit ourself to stress that
naked singularities exist in self-similar solutions
with pressure, thereby showing that the phenomenon
of naked singularities formation cannot be considered as an artifact
of dust (i.e. vanishing stresses) solutions.

\subsubsection{General stresses}

The problem of predicting the final fate of an initially regular distribution
of matter supported
by an arbitrary distribution of stress-energy
(including e.g. the case of isotropic perfect fluids but also
anisotropic crystalline structures which are thought to form at extremely
high densities), is still open even in the spherically symmetric case.
First of all, one has to take into account the fact that
there is a high degree of uncertainty
in the properties of the equation of state at very high
densities. Recently, Christodoulou~\cite{ChristodoulouLNP}
initiated the analysis of
a simple model composed by a dust (``soft'') phase 
for energy density below a certain
value $\overline{\epsilon}$ and a stiff (``hard'') phase
for $\epsilon>\overline{\epsilon}$ (in the hard phase the pressure
is given by $p=\epsilon -\overline\epsilon$).

Although the details of the collapse in presence of general matter fields 
are still largely unknown, 
it is very unlikely that
the ``embarrassing'' examples of naked singularity formation like 
those occurring in dust
can eventually be eliminated
with the ``addition'' of stresses.
Indeed, it is reasonable to think that a sector of naked singularities 
exists in the choice of initial data for any fixed equation of 
state~\cite{JoshiCMP}.
The main issues that have, therefore, to be addressed are the {\it genericity}
and the {\it stability} of naked singularity formation.

Both the above italicized terms 
have a somewhat intuitive meaning that is, however, difficult
to express in mathematical terms.
Regarding genericity, one can mean that the set of initial data
leading to naked singularities is not of measure zero.
For instance, it has been shown~\cite{Jhingan}
that among the dust solutions $s = \{F,f,1\}$
naked singularities are generic in the sense that at a fixed density 
profile $F$, one can always choose energy functions
$f$ leading to blackholes or naked singularities.
A generalization of such a result would be that the naked singularities
are generic - in this specific sense - at fixed, but arbitrary, equation of 
state $\psi$ (there is some convincing evidence for this, 
see~\cite{JoshiCQG}).

The issue of stability is even more delicate than that of genericity.
Indeed, any exact solution of a physical theory
must survive to small but arbitrary perturbations in order to
serve as a candidate for describing nature.
In the case of exact solutions of the Einstein field equations, 
the notion of stability is very delicate due to 
to the gauge invariance of the theory
with respect to spacetime diffeomorphisms. 
Recently, some evidence of stability of dust naked singularities against
perturbations has been obtained~\cite{Harada1}. 

\section{Discussion: a picnic on the side of the road}

It is well known that, if the mass of a collapsing object does not
fall below the neutron star limit ($\sim 3M_\odot$), no physical
process is able to produce enough pressure to balance the gravitational pull
so that continued gravitational collapse must occur. It is widely 
believed that the final state of this process is a blackhole.
However, what general relativity actually
predicts, in the cases which have been analysed so far,
is that either 
a blackhole or a naked singularity is formed, depending on the 
initial distribution of  density and velocity and on the constitutive
nature of the collapsing matter. 
One may raise the objection
that most of the known analytical results 
could be an artifact of spherical symmetry. However, from the numerical
point of view some evidence that this is not the case is coming up. 
Therefore, as singularity theorems showed that the singularities 
occurring in collapse are generic and not any artifact of
symmetry, a similar situation may hold for the nature of singularities
as well.

One may at this point ask if and when, a cosmic censorship  theorem holds in
nature. An answer to such a question could be in the negative. However,
it remains to understand  the physics underlying 
the end state of gravitational collapse with respect to the choice of initial 
data at  fixed matter model.
In a famous novel~\cite{strug} (that inspired the film {\it Stalker}
by A. Tarkovsky) a short visit of extraterrestrial life on the earth occurs.
The gap between the two civilizations is so high that 
human beings are, with respect to the ``garbage'' left by the visitors,
like ants exploring what remains on the side of the road
after a human picnic. Something they find is
useful, something useless, something even dangerous,
but anyway everything is obscure and difficult to understand,
looking like the weak shadow of a wonderful abyss of knowledge.
Our present cosmic censorship understanding resembles this situation.
Indeed, we are getting a variety of mathematical
hints with somewhat obscure physical meanings. For instance: 
the condition on $\xi$ recalled in Section 4.2.4,
the constraints arising in the gravitational collapse of Einstein clusters,
the dimensionless numbers arising in Choptuik's
numerical results~\cite{Gundlacrev}, the condition on
the radiation flux which arises in Vaidya collapse~\cite{Global}.
Such ``numbers'' should presumably be the remnant
weak shadow of a general theorem when its hypotheses are enormously 
restricted by the choice of the equation of state and of the adopted 
symmetries.

To get rid of this puzzle appears to be one of the
most exciting objectives of future research in classical relativity theory.   

\section*{Acknowledgements} 
Many discussions with Elisa Brinis and Pankaj Joshi are gratefully 
acknowledged. S. J. thanks the ICSC World Laboratory (Lausanne, Switzerland)
for the Chandrasekhar Memorial Fellowship (1998-99).

\end{document}